\documentclass[10pt, twocolumn,aps,pra,superscriptaddress]{revtex4-2}

\usepackage{graphicx}
\usepackage{textgreek}

\usepackage[] { hyperref } 
\hypersetup{
    colorlinks=True,
    linkcolor=purple,
    citecolor=blue,   
    urlcolor=blue,
    }
\usepackage[utf8]{inputenc}
\usepackage[T1]{fontenc}
\usepackage{tabularx}
\usepackage{svg}
\usepackage[normalem]{ulem}
\usepackage[dvipsnames]{xcolor}
\usepackage{titlesec}

\usepackage[uncertainty-mode = separate, separate-uncertainty-units = single,  bracket-unit-denominator = false]{siunitx} 
\AtBeginDocument{\RenewCommandCopy\qty\SI} 

\titleformat{\section}
 {\bfseries\fontsize{11pt}{16pt}\selectfont}     
  {\thesection}    
{}            
  {\centering}
  
\titlespacing{\section}
 {0pt}   
 {0pt}   
 {10pt}

\newcommand{\e}{\mathrm{e}}

\renewcommand{\epsilon}{\varepsilon}

\newcommand{\cnn}{Universit\'e Paris-Saclay, Centre de Nanosciences et de Nanotechnologies, CNRS, 10 Boulevard Thomas Gobert, 91120 Palaiseau, France}
\newcommand{\quandela}{Quandela, 7 Rue Léonard de Vinci, 91300 Massy, France}

\begin{document}

\title{Near-identical photons from distant quantum dot-cavity devices}
\author{Thibaut Pollet}
\affiliation{\cnn} 
\author{Victor Guilloux}
\affiliation{\cnn}
\author{Duc-Duy Tran}
\affiliation{\quandela}
\author{Anton Pishchagin}
\affiliation{\quandela}
\author{Stephen Wein}
\affiliation{\quandela}
\author{Joseph A. Sulpizio}
\affiliation{\quandela}
\author{William Hease}
\affiliation{\quandela}
\author{Petr Stepanov}
\affiliation{\quandela}
\author{Petr Steindl}
\affiliation{\cnn}
\author{Nico Margaria}
\affiliation{\quandela}
\author{Samuel Mister}
\affiliation{\quandela}
\author{Martina Morassi}
\affiliation{\cnn}
\author{Aristide Lemaître}
\affiliation{\cnn}
\author{Thi Huong Au}
\affiliation{\quandela}
\author{Sébastien Boissier}
\affiliation{\quandela}
\author{Pascale Senellart}
\affiliation{\cnn}

\date{\today}

\begin{abstract}
Scalable optical quantum technologies require interference between large numbers of indistinguishable single-photons emitted by independent sources. Semiconductor quantum dots are known to be excellent on-demand sources of single-photons. They show record efficiency when inserted into optical cavities to control their spontaneous emission and generate trains of near identical photons over  microsecond timescales. However, generating perfectly identical photons from distant cavity-based sources has remained a long-standing challenge. It requires precise matching of the emission wavelengths and emission dynamics, while simultaneously minimizing spectral noise across all time scales for distant emitters in uncorrelated environments. Here, we report on the nanofabrication of a large number of quantum dot-cavity sources with ultra-low spectral noise and wavelength dispersion. The high source efficiency and the use of two tuning mechanisms enable precise optimization of the spectral overlap between distant sources. With this approach, we demonstrate a two-photon indistinguishability of \qty{88\pm1}{\percent} between photons emitted from two distant sources. Remarkably, this value reaches the upper bound set by the intrinsic indistinguishability of photons emitted successively by each source. These results represent a key milestone for scaling photon-based quantum technologies.
\end{abstract}
\maketitle

Optically active epitaxial quantum dots (QDs) can serve as deterministic sources of single photons \cite{senellart_high-performance_2017}, making them an essential building block for quantum networks \cite{kimble_quantum_2008, beccaceci_all-photonic_2025} and photonic quantum computing \cite{wein_minimizing_2025, gliniasty_spin-optical_2024, dessertaine_enhanced_2026, chan_tailoring_2025, chan_practical_2025}. For large-scale applications, the simultaneous generation of many single-photons requires an emission and collection efficiency close to unity. This can be achieved by coupling the QD to a cavity, which accelerates the emission and efficiently funnels it into an output mode \cite{somaschi_near-optimal_2016, wang_micropillar_2020, arcari_near-unity_2014, ding_high-efficiency_2025}. Because photons do not interact directly, entangling operations rely on quantum interference and measurement. Such interference is maximized when the photons are identical in all degrees of freedom that do not encode the quantum information, so that no unintended distinguishing information is revealed by the measurement. The Hong–Ou–Mandel (HOM) interference 
\cite{hong_measurement_1987} is the canonical signature of this indistinguishability, which underpins photonic two-qubit gates \cite{kok_linear_2007}, entangling measurements \cite{browne_resource-efficient_2005}, and remote entanglement generation \cite{delteil_generation_2016}. The emission of indistinguishable photons from a single QD-based source, with emission time separations ranging from a few nanoseconds to several microseconds, is now well-established \cite{somaschi_near-optimal_2016,ding_high-efficiency_2025,margaria_efficient_2025,tomm_bright_2021}. In parallel, impressive progress in the emission of indistinguishable photons from remote QDs has been achieved with droplet-etched GaAs QDs in bulk, i.e. at low emission rates and with phonon side-band filtering \cite{zhai_quantum_2022}. Achieving high indistinguishability for efficient cavity-based single-photon sources requires overcoming additional challenges \cite{gold_two-photon_2014, reindl_phonon-assisted_2017, thoma_two-photon_2017, you_quantum_2021, strobel_telecom-wavelength_2025, laneve_quantum_2025, pont_indistinguishability_2025}. First, the QDs must have matching emission wavelengths and lifetimes, with the latter strongly depending on the photonic environment in which they are embedded through the Purcell effect. Precise control over multi-step fabrication processes is thus needed to achieve highly reproducible photonic properties. Second, multi-step fabrication processes and surface proximity have been shown to create additional source of noise \cite{wang_optical_2004, liu_single_2018}. Lastly, each QD experiences different noise through the coupling to their respective solid-state environments, leading to spectral fluctuations across different timescales \cite{vural_perspective_2020, kuhlmann_charge_2013, kambs_limitations_2018, dangel_two-photon_2022}. 
In this paper, we demonstrate the deterministic fabrication of near-identical micropillar cavities embedding emitters grown using a method that ensures very low spectral noise. We exploit two independent tuning mechanisms to control the QD emission wavelength, one based on electric fields and the other on applied strain. Using this method, we achieve a two-photon indistinguishability as high as \qty{88\pm1}{\percent} for remote sources, a record value limited only by the residual dephasing taking place at nanosecond timescales for each individual emitter. All noise sources at longer time scales are shown to be negligible. \\  

\begin{figure*}[ht]
    \includegraphics[width=17.2cm]{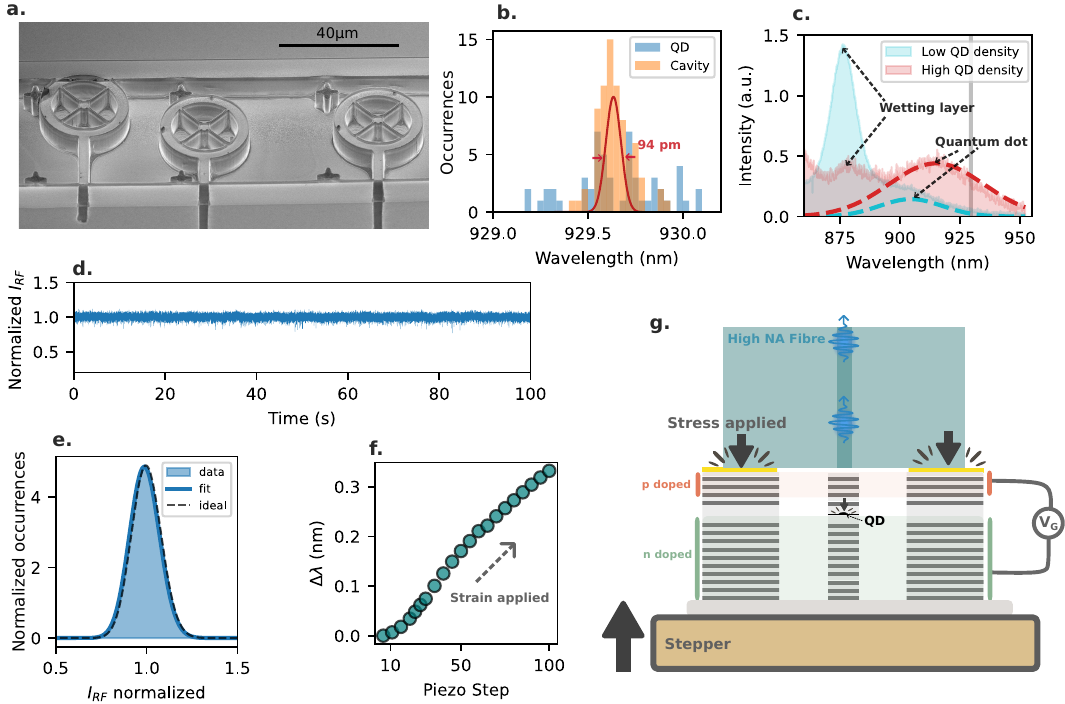}
    \caption{\label{fig1} \textbf{Fabrication of low-noise identical single-photon sources.} \textbf{a.} Scanning electron microscope image of three micropillar single-photon sources.
    \textbf{b.} Wavelength distribution of fabricated micropillar cavities, fitted to a normal distribution yielding a FWHM of \qty{94(2)}{\pico\meter}. The wavelength distribution of the QDs embedded in the micropillars shows a larger standard deviation requiring post-fabrication tuning to bring the QDs into resonance with their respective cavities.
    \textbf{c.} Spectral distribution of the QD ensemble under above-band excitation for two wafers with different QD areal densities, fitted with Gaussian functions. The pronounced wetting-layer emission in the blue curve indicates that QD growth is terminated at the onset of dot formation, resulting in a lower QD areal density. This low-density wafer exhibits reduced charge noise and is therefore used in this work \cite{pollet_et_al_low-noise_nodate}. The grey shaded area shows the spectral window for the QD selection.
    \textbf{d.} Time trace of the RF intensity $I_\mathrm{RF}$ normalized by its mean value for a QD embedded in a micropillar (S1) using CW excitation (shown with a binning of \qty{1}{\milli\second}). The laser is significantly narrower and more stable than the QD, so intensity variations directly reflect fluctuations in the QD emission wavelength
    \textbf{e.} Histogram of the RF time trace (binning of \qty{100}{\micro\second}), fitted with a sum of Poissonian distributions weighted by a Gaussian distribution (Methods).
    A Gaussian distribution with $\sigma_\text{noise}= \qty{0.11}{\pico\metre}$ fits the spectral fluctuation induced by charge noise \cite{matthiesen_full_2014, berthelot_random_2010}. 
    The Poissonian distribution, plotted in a dashed line, represents the shot-noise limit. \textbf{f.} QD wavelength shift versus the number of steps applied to the piezo stepper on which the sample is mounted. Each step increases the strain applied to the QD, resulting in tuning of its emission wavelength. \textbf{g.} Schematic view of the fibre-based device. A high NA (0.35) fibre is placed on top of the micropillar and makes contact with the sample contact pads. The sample is pressed against the fibre by a stepper piezo placed below it, applying stress that propagates to the QD and shifting its emission wavelength. The sample is electrically connected to allow control of the QD charge state.}
\end{figure*}

\begin{figure*}[ht]
    \includegraphics[width=17.2cm]{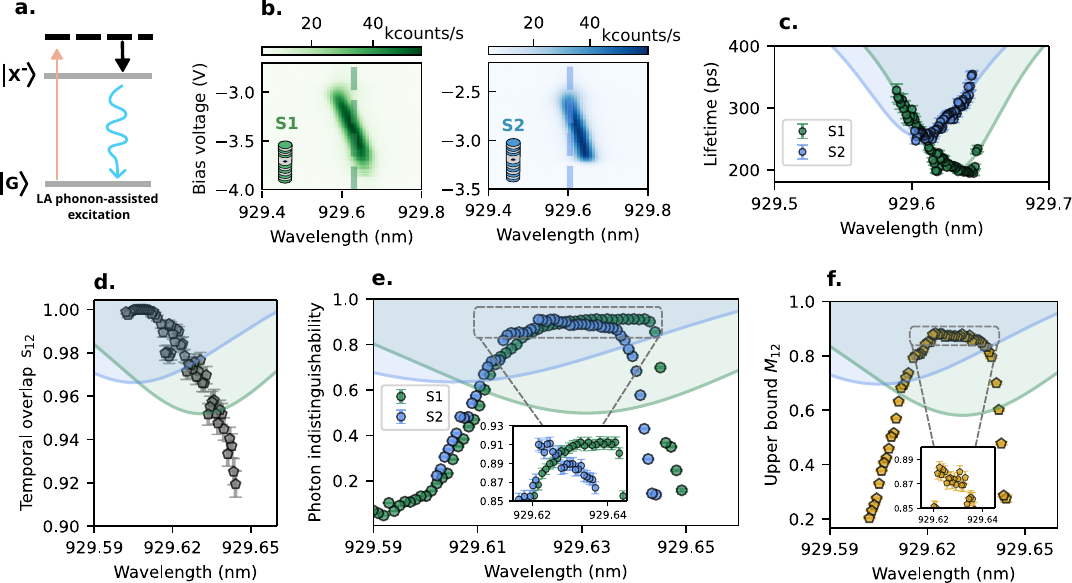}
    \caption{\label{fig2} \textbf{Individual performance of the two QD-cavity single-photon sources.} \textbf{a.} Working principle of the LA phonon-assisted excitation scheme.
    \textbf{b.} Charge plateau of the negative trion $X^-$ state for the two QDs embedded in the separate micropillars S1 and S2. The dashed lines represent the cavity resonance of each micropillar.
    \textbf{c.} Lifetimes of S1 and S2 measured at different wavelengths by varying the PIN bias. The data points are fitted with a Lorentzian function to model the detuning-dependent Purcell effect. The fitted curve is reproduced in the following three figures to help visualize the effect of the cavities.
    \textbf{d.} Temporal overlap calculated using Eq.\,\ref{eq1} from the extracted lifetimes of the two emitters. \textbf{e.} Indistinguishability of successively emitted photons measured individually for S1 and S2 with an unbalanced Mach-Zehnder interferometer (Methods). The inset displays a zoom where the indistinguishability is maximal.
    \textbf{f.} Expected upper bound for $M_{12}$ extracted from the individual source properties using Eq.\,\ref{eq2}. Again, the inset provides a zoom on maximal values.
    }  
\end{figure*}

\section*{Fabrication of low-noise identical single-photon sources}
\indent \indent Our single-photon sources are based on individual InGaAs QDs embedded in electrically connected micropillar cavities \cite{somaschi_near-optimal_2016, ollivier_reproducibility_2020, margaria_efficient_2025}. The cavity heterostructure forms a PIN diode to control the charge state of the QDs and to fine-tune the QD emission wavelength (Fig.\,\ref{fig1}a, Methods for the sample design). QDs are selected within a narrow spectral window centered on the cavity target wavelength, and micropillars are fabricated around individual QDs using in-situ lithography followed by etching \cite{dousse_controlled_2008}.
We achieve a high level of precision on the fabricated cavity wavelength by compensating the wafer epitaxial thickness gradient (\qty[per-mode = symbol]{0.16}{\nm\per\mm}) with slight individual adjustments of the micropillar diameters (Extended Data Fig.\,\ref{extfig1}). Fig.\,\ref{fig1}b shows the cavity wavelength distribution of \num{75} micropillars from the sample used in this work. A fit to a normal distribution yields a full width at half maximum (FWHM) of \qty{94(2)}{\pico\meter}, a value that lies within the average cavity mode linewidth (\qty{116}{pm}). The wavelength distribution of the QDs embedded in the micropillars shows a FWHM of \qty{355}{pm}. This distribution is mostly determined by the spectral window used during QD selection in the in-situ lithography step and represents only a minor issue when using post-fabrication tuning techniques, as described below. The selection window can be narrowed, at the expense of an increased selection time for suitable QDs.

A limiting factor for the emission of indistinguishable photons from remote sources is the uncorrelated noise arising from the local solid-state environment. By using high-purity epitaxial growth that yields a low areal density of QDs on the wafer (see Fig.\,\ref{fig1}c), we limit the presence of charge traps and thus largely suppress charge noise \cite{pollet_et_al_low-noise_nodate}. 
Furthermore, the PIN diode enables efficient removal of fluctuating charge carriers under reverse bias operation. By using continuous-wave (CW) resonant excitation \cite{steindl_cross-polarization-extinction_2023}, we observe stable emission from the emitters, as illustrated by the resonance fluorescence (RF) time trace shown in Fig.\,\ref{fig1}d. The corresponding intensity distribution (Fig.\,\ref{fig1}e) shows only a small deviation from the ideal shot-noise-limited Poissonian distribution, indicating a minimal contribution of charge noise to the emitter linewidth.
A fit to the distribution gives an average spectral fluctuation value of \qty{4.8}{\percent} of the natural linewidth. For emitters with identical lifetimes and lifetime-limited linewidths, we estimate an achievable mutual indistinguishability value as high as \qty{99.5}{\percent} for this amount of spectral noise (Methods).

To address the residual variation in the QD emission wavelength after micropillar etching, we implement two complementary tuning mechanisms based, respectively, on the application of electric and strain fields. The application of a bias voltage to the PIN diode allows control of the QD charge state by Coulomb blockade and fine-tuning of the QD emission wavelength by a vertical electric field \cite{warburton_single_2013}.
In this work, we exclusively use the negative trion state $X^-$ which has a simpler decay temporal profile than the neutral exciton state \cite{ollivier_reproducibility_2020}.
The tuning range obtained by this method is limited by the extent of the Coulomb plateau, measured to be approximately \qty{50}{\pm} wide (Fig.\,\ref{fig2}b). This is insufficient to fully compensate for the relative detuning of two arbitrary QDs given their wavelength distribution. To increase the tuning range, we implement strain tuning of the QD emission by applying localized stress on the sample with an optical fibre. We use a high Numerical Aperture (NA) fibre placed on top of the device (Fig.\,\ref{fig1}g), which is well-coupled to the micropillar cavity to enable efficient collection of the emitted single photons \cite{margaria_efficient_2025}.
After bringing the fibre in contact with the sample, we apply stress by pressing the fibre against the sample using a z-axis piezo stepper. The stress propagates from the fibre to the contact pads surrounding the micropillar cavity and then to the QD inside the micropillar via the connecting arms, resulting in a shift of the QD wavelength from strain. This provides a complementary tuning range of hundreds of picometers (Fig.\,\ref{fig1}f). The wavelength of an arbitrary QD can thus be coarsely tuned by applying strain with the fibre, 
while the bias voltage is mainly used to control the QD charge state and for fine wavelength adjustment.\\ \\
\section*{Individual source performance and upper bound}
\begin{figure*}[ht]
    \includegraphics[width=18cm]{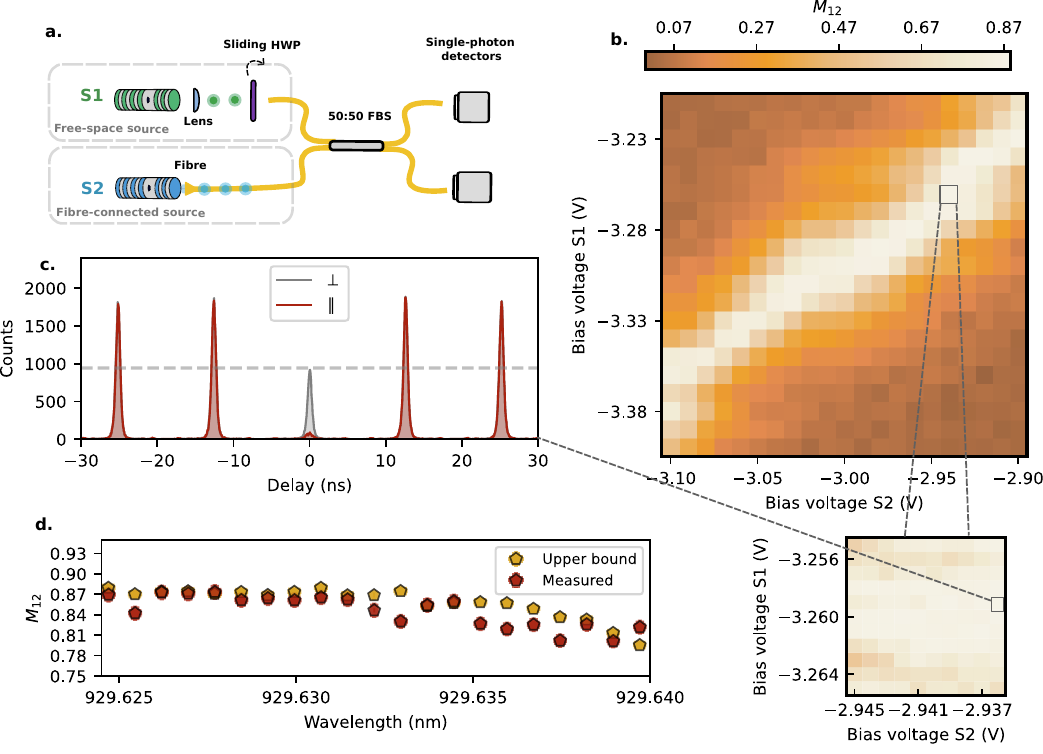}
    \caption{\label{fig3} \textbf{Two-photon interference between remote QDs embedded in two separate micropillar cavities.} \textbf{a.} Illustration of the HOM interference setup, including the motorized HWP placed in one of the optical paths to switch between orthogonal ($\perp$) and parallel ($\parallel$) polarization configurations (see Extended Data Fig.\,\ref{extfig3} for more information). \textbf{b.} Remote-source indistinguishability as a function of emitter wavelengths, measured by varying the PIN bias voltage of each sources. At each pair of wavelengths, the photon correlations in the $\perp$ and $\parallel$ configurations are recorded for \qty{1}{\second}. We plot the visibility extracted using Eq.\,\ref{eq3} and corrected for setup imperfections (Methods).
    We perform a finer scan around the maximum of the visibility, as shown in the bottom part. \textbf{c.} HOM interference from photons emitted by two distant single-photon sources, showing a corrected mutual indistinguishability  of $(88\pm1)\%$. The histogram represents the maximum HOM visibility obtained by varying the voltages of both sources. The grey-dashed line shows the limit when the photons are distinguishable ($\perp$ case) and do not interfere.
    \textbf{d.} Comparison of the upper bound and the measured mutual indistinguishability as a function of the emission wavelength and in the absence of spectral detuning between S1 and S2.}
\end{figure*}
\indent\indent Leveraging precise cavity fabrication, low-noise emitters, and emission wavelength control, we fabricate a large number of micropillars on one microchip, all designed to target the same cavity wavelength (Fig.\,\ref{fig1}a). To avoid variations which might arise from processing different wafers, we cleave our chip into two pieces, each containing approximately 40 micropillars.
Both pieces are placed on two piezo stacks inside a single closed-cycle cryostat and cooled down to \qty{5}{\kelvin}. One sample (denoted S1) is placed under a lens for free-space optical collection of the emitted photons. The other sample (denoted S2) is mounted under an optical fibre for both photon collection and strain tuning of the QD emission line (Fig.\,\ref{fig1}g).
The PIN diode on each sample is connected to an ultra-low-noise voltage source, enabling independent bias voltage tuning. To trigger emission from the QD, we use the longitudinal acoustic (LA) phonon-assisted excitation scheme with \qty{15}{\ps} short laser pulses, blue-detuned by \qty{0.8}{\nm} from the $X^-$ state wavelength (Fig.\,\ref{fig2}a) \cite{thomas_bright_2021}. The residual laser is suppressed using three \qty{0.8}{\nm} wide band-pass filters which transmit the QD emission line.
We first select a micropillar on S1 whose QD is closest to resonance with the cavity. Next, we select the micropillar cavity on S2 whose resonance is closest to that of the selected cavity on S1. The two selected cavities are detuned only by \qty{25}{\pico\meter}.
Finally, we tune the QD on S2 to the same wavelength as the QD on S1 using the fibre-based strain-tuning method described above. We ensure that the charge plateaus of both trions overlap (Fig.\,\ref{fig2}b) to maximize the achievable tuning range around the optimum operating point obtained by applying electrical biases to the PIN diodes. To achieve this, it was necessary to slightly detune the QD on S2 from its cavity resonance (Fig.\,\ref{fig2}c).

Maximizing the indistinguishability requires both fine-tuning of the emission wavelength as well as matching of the QD lifetimes.
The lifetime of each source varies along the trion charge plateau (see Fig.\,\ref{fig2}d) with a significant reduction of the radiative lifetime by Purcell enhancement from the cavity resonance.
Because of variation in the bulk decay rates and achieved Purcell factors, together with the presence of a small detuning between the two cavities, the QDs do not reach perfectly identical lifetimes for the same wavelengths.
Hence, we observe (as shown in Fig.\,\ref{fig2}d) a small wavelength-dependent temporal overlap $s_{12}$ between the two single-photon wavepackets, as defined by
\begin{equation}
    s_{12}=\frac{4\tau_1\tau_2}{(\tau_1+\tau_2)^2}\,,
    \label{eq1}
\end{equation}
where each individual QD can be characterized by a mono-exponential decay given by the trion emission lifetime $\tau_1$ ($\tau_2$) for S1 (S2) \cite{pont_indistinguishability_2025}. Within the wavelength range of the charge plateau, the temporal overlap remains higher than \qty{0.92\pm0.05}{} and reaches near unity around \qty{929.61}{\nm}.
With the QDs set at the middle of their charge plateaus, polarized single-photon count rates of \qty{3.1}{\mega\hertz} for S1 and \qty{1.5}{\mega\hertz} for S2 are measured at the output of the laser filtering stages using an avalanche photodiode (APD). Considering the \qty{79.3}{\mega\hertz} laser repetition rate and accounting for the detector efficiency ( $\approx$ \qty{31}{\percent} at \qty{929}{\nm}) and the setup transmission, we infer a probability of collecting a polarized photon at the output of the micropillar of \qty{27.5}{\percent} (\qty{15.6}{\percent}) for S1 (S2). Note that the two fundamental cavity modes are degenerate, so not to polarize the emission and preserve the selection rules between the spin and the polarization of the emitted photons for future applications. For the $X^-$ state, we expect a factor \num{2} increase for the unpolarized brightness, yielding \qty{55}{\percent} (\qty{31.2}{\percent}) for S1 (S2).

Remote source interference also depends on the indistinguishability of sequentially emitted photons from the individual sources, which is affected by dephasing occurring on timescales comparable to the lifetime or within the \qty{12}{\nano\second} interval between successive photons. We first note that under LA excitation, S1 (S2) exhibit a high single-photon purity with $g^{(2)}(0) = \qty{1.4\pm.3}{\percent}$ ($g^{(2)}(0) = \qty{2\pm0.4}{\percent}$) (Extended Data Fig.\,\ref{extfig2}).
In Fig.\,\ref{fig2}e, we show the indistinguishability of consecutively emitted photons for both sources as a function of the bias-controlled emission wavelength. For S1 (S2), we achieve a maximum corrected individual indistinguishability value of $M_{1,\textrm{max}}=\SI{91.3\pm0.7}{\percent}$ ($M_{2,\textrm{max}}=\SI{91.1\pm0.8}{\percent}$) (Methods), where the high indistinguishability is obtained by operating the source at the center of the trion charge plateau. Matching emission wavelengths and lifetimes alone is not sufficient for high mutual indistinguishability. Each source must also have a stable charge state, ideally by operating away from the edges of the charge plateau. This requirement underscores the importance of having a diode structure for QD-based single-photon sources.
We note that the sources do not reach their highest individual indistinguishabilities at the same wavelength due to a small dependence on the Purcell-enhanced emitter lifetime. This adds another constraint to the determination of the optimal operating point beyond the temporal overlap.
Knowing the wavepacket overlap $s_{12}$ and the individual indistinguishabilities $M_i$ of the two sources, we can infer an upper bound on the mutual indistinguishability between the remote sources $M_{12}$ when there is no spectral detuning between S1 and S2:
\begin{equation}
    M_{12}\le \frac{2s_{12}M_1M_2}{(M_1+M_2)+(M_1-M_2)r}\,,
    \label{eq2}
\end{equation}

where $r=\pm\sqrt{1-s_{12}}$, and we take $r \geq 0$ when $\tau_1 \leq \tau_2$ (see Supplementary for the derivation).
The calculated upper bound is plotted in Fig.\,\ref{fig2}f as a function of the emission wavelength, reaching a maximum value of $M_{12}=\qty{88.3\pm0.5}{\percent}$.\\

\section*{Two-photon interference from distant QD-cavity devices}

\indent \indent Finally, we perform an HOM interference measurement between photons emitted by S1 and S2.
As sketched in Fig.\,\ref{fig3}a, photons from both sources are sent to the two input ports of a \num{50}:\num{50} fibre beam splitter (FBS), and two-photon coincidence counts are recorded at its outputs using single-photon detectors (Methods).
We use the same pulsed laser to simultaneously trigger both single-photon sources (Extended Data Fig.\,\ref{extfig3}). To compensate for the difference in brightness between the two sources, we attenuate the brighter source (S1) to match the count rate of the second one (S2). 
Crucially, we perform the experiment without rejecting any useful photons: no narrow spectral filtering or temporal post-selection is used. No active stabilization of the QD emission wavelength is employed either. By avoiding optical losses due to filtering, we are thus able to fully exploit the brightness of the sources and perform correlation measurements with a raw average count rate of \qty{1.5}{\mega\hertz} per detector, and a raw two-photon coincidence rate of \qty{28}{\kilo\hertz}.

To maximize photon indistinguishability between the remote sources, we scan the source wavelengths by varying the bias voltages of S1 and S2, and record the intensity correlation histogram at each point. A half-wave plate (HWP) is mounted on a motorized slider, so it can be inserted into the optical path to rotate the polarization of S1 from parallel ($\parallel$) to orthogonal ($\perp$) relative to S2 (Extended Data Fig.\,\ref{extfig4}). In total, we record histograms in the parallel and orthogonal configurations for \qty{420}{} sets of voltage parameters (the complete scan took just under \qty{1}{\hour} of run time). The coincidences in the central peaks are integrated using a \qty{4}{\ns} time window and we extract the mutual indistinguishability $M_{12}$ (Methods) between single-photons emitted by S1 and S2 (Fig.\,\ref{fig3}b).
To discard artifacts such as count-rate imbalance or blinking \cite{jons_two-photon_2017, weber_overcoming_2018}, we verify that the zero-delay peak measured in the perpendicular configuration is exactly one half of the coincidences recorded with non-zero delays. We also confirm that long-delay correlations were not subject to exponential decay (Extended Data Fig.\,\ref{extfig5}).

In the absence of wavelength detuning between the two emitters, we observe a high degree of mutual indistinguishability, which is maintained for different wavelengths across the charge plateaus. To find the highest HOM visibility, we scan the bias voltages of both S1 and S2 using an iterative grid refinement strategy with \qty{10}{\milli\volt} and \qty{1}{\milli\volt} step sizes, respectively. The highest HOM visibility of $V_\textrm{rem}=\qty{85.3\pm1}{\percent}$ is found when both sources emit at \qty{929.629}{\nm}, corresponding to a mutual indistinguishability of $M_{12}=\qty{88\pm1}{\percent}$ (Fig.\,\ref{fig3}c), corrected for the imbalance of the FBS and the non-zero $g^{(2)}(0)$ of the sources. To the best of our knowledge, this sets a new record for a two-photon interference between remote solid-state emitters embedded in cavities. 
In Fig.\,\ref{fig3}d, we compare the indistinguishability measured at different wavelengths, in the absence of emitter detuning, to the theoretical upper bounds given by Eq.\,\ref{eq2}. We find that the measured values are in close agreement with the calculated upper bounds across the full range of \qty{15}{\pm}.
This highlights the high stability of our single-photon sources, as charge noise does not noticeably degrade $M_{12}$. We note again that this result is achieved without narrow spectral filtering of the single-photon emission.

Our measurements show that the uncorrelated noise between the two independent sources is effectively eliminated, and that the mutual photon indistinguishability is mostly limited by the intrinsic indistinguishability of consecutive photons emitted by each source.
The upper bound could therefore be further improved with better single-source indistinguishability limited by phonon-induced pure dephasing and side-band emission \cite{reigue_probing_2017}. Reducing the emitter lifetimes with a stronger Purcell effect would enhance the single source indistinguishability visibility \cite{grange_reducing_2017}, which can be achieved by increasing the cavity quality factor. In this work, the devices had an average quality factor of \qty{8000}{} as a compromise between optimal source performance and the yield of spectrally identical cavities. By leveraging micropillar devices with $Q=\qty{13000}{}$ showing state-of-the-art indistinguishability  \cite{margaria_efficient_2025}, and assuming perfect temporal overlap $s_{12}=\qty{1}{}$, we expect an achievable upper bound of $M_{12}=\qty{97.5}{\percent}$ in the near-term.\\ \\

\section*{Conclusion}
\indent\indent In conclusion, we have reported on an unprecedented degree of control in the fabrication of single-photon sources based on InGaAs QDs embedded in cavities. Owing to high-purity epitaxial growth and low QD density, negligible QD spectral noise is observed. By combining two QD tuning mechanisms, a mutual indistinguishability as high as \qty{88\pm1}{\percent} is reported for distant cavity-based emitters without spectral filtering and at high emission rates. This value is mostly limited by residual pure dephasing of the individual emitters induced by phonons. The reported fabrication technique for identical single-photon sources is readily scalable and can be easily automated. Our results pave the way for integrating multiple independent single-photon sources into current photonic quantum computers, eliminating the need for complex demultiplexing schemes and increasing sampling rates \cite{maring_versatile_2024}. Furthermore, our measurements were performed  with negatively charged trions under phonon-assisted excitation, an experimental configuration that allows the deterministic generation of reconfigurable spin-photon cluster states \cite{coste_high-rate_2023, huet_deterministic_2025}. Finally, our results also constitute a fundamental building block toward fault tolerant quantum computation with the spin-optical quantum computing architectures, by enabling photon-mediated spin–spin entanglement \cite{wein_minimizing_2025, gliniasty_spin-optical_2024, chan_tailoring_2025, chan_practical_2025}.

\bibliography{references}

\newpage
\section*{Methods}
\noindent
\textbf{Sample design.}
Our single-photon sources are based on low-density, self-assembled InGaAs QDs, grown by molecular beam epitaxy in the Stranski–Krastanov growth mode. The QD layer is deterministically embedded at the field antinode of a $\lambda$-sized planar cavity formed by two distributed Bragg reflectors (DBRs) with \num{18} top and \num{36} bottom pairs of AlGaAs/GaAs $\lambda/4$ layers. The QDs are pre-selected and marked using the in-situ lithography technique \cite{dousse_controlled_2008}. The semiconductor material surrounding the QDs is then etched to define micropillar structures as shown in Fig.\,\ref{eq1}a \cite{somaschi_near-optimal_2016, margaria_efficient_2025}.
To form a PIN diode structure, part of the top mirror is p-doped with carbon acceptors, while the bottom mirror is n-doped with silicon donors. To bias the QDs, the micropillars are electrically connected through four contact arms to a larger diode structure for wire bonding.\\ \\
\textbf{Spectral fluctuations under resonant excitation.} To quantify the stability of the single-photon sources, we implement an analysis method based on the time trace of the resonant fluorescence (RF) intensity, as introduced in Ref. \cite{matthiesen_full_2014}. For sufficiently large time bins to avoid the antibunching dip, the temporal intensity distribution for an ideal emitter under continuous-wave resonant excitation follows Poisson statistics. Spectral fluctuations of the transition energy reduces the emission intensity due to detuning from the excitation laser, causing the temporal distribution to deviate from an ideal Poissonian distribution. We model this deviation as a sum of Poissonian distributions with lower mean values relative to the maximum mean intensity value. By doing so, we directly link the intensity distribution to energy fluctuations induced by charge noise. 

We model the spectral fluctuations with a stationary process where the energy detuning $\Delta$ follows Gaussian statistics at any given time \cite{berthelot_random_2010}
\begin{equation}
W(\Delta) = \frac{1}{\sqrt{2 \pi \sigma_\text{noise}^2}} \, \exp\Bigg( -\frac{(\Delta - \mu)^2}{2 \sigma_\mathrm{noise}^2} \Bigg)\,,
\end{equation}
with $\mu$ the mean value of the detuning and $\sigma_\mathrm{noise}$ the standard deviation.
We fit the intensity distribution with a Gaussian-weighted sum of Poissonian distributions (Fig.\,\ref{fig1}e):
\begin{equation}
P(k) = \int W(\Delta)\frac{m(\Delta)^k\exp(-m(\Delta))}{k!}\,d\Delta\,,
\end{equation}
where $k$ is the number of single-photon events per time bin of $t_\text{bin}=\qty{100}{\micro\s}$ and $m$ is the detuning-dependent mean value of the Poissonian distribution which has a Lorentizan profile. 
From Fig.\,\ref{fig1}(e), we find $\sigma_\mathrm{noise}=\qty{0.16}{\micro\e\volt}$ or \qty{0.11}{\pico\meter}; corresponding to \qty{4.8}{\percent} of the natural linewidth ($\sigma_\mathrm{noise}=0.048/\tau$). \\ \\
\textbf{Bounds on remote source interference.} The mutual indistinguishability $M_{12}$ between photons emitted by two exponentially-decaying emitters with lifetimes $\tau_1$ and $\tau_2$ is given by \cite{wein_analyzing_2020, pont_indistinguishability_2025}:
\begin{equation}
    M_{12} = s_{12}\frac{(\Gamma_1+\Gamma_2)(\gamma_1+\gamma_2)}{(\Gamma_1+\Gamma_2)^2+4\Delta^2},
\end{equation}
where $\gamma_i=1/\tau_i$ is the $ i^{\text{th}} $ emitter decay rate, $\Gamma_i = \gamma_i/M_i$ is the decay rate of the coherence visibility corresponding to a source indistinguishability $M_i$, and $s_{12}=4\gamma_1\gamma_2/(\gamma_1+\gamma_2)^2$ is the temporal mode overlap when $M_1=M_2=1$. Setting $\Delta=0$ and expressing the result in terms of $\tau_i$ and $M_i$, we obtain
\begin{equation}
    M_{12} \leq M^\prime_{12} = s_{12}\frac{(\tau_1+\tau_2)M_1M_2}{\tau_1M_1+\tau_2M_2},
\end{equation}
where $M^\prime_{12}$ is an upper bound on the remote source indistinguishability set by the individual source indistinguishabilities $M_1$ and $M_2$, and any mismatch in their lifetimes $\tau_1$ and $\tau_2$, which yields $s_{12}\leq 1$. By simplifying this result, we recover an expression for the upper bound which depends on $s_{12}$ only, as given in Eq.\,\ref{eq2} of the main text.\\ \\ \textbf{Impact of charge noise on remote source interference.}
To assess the impact of charge noise on the remote source interference visibility, we take the mutual indistinguishability $M_{12}$ and average it over a detuning $\Delta$ sampled from a Gaussian distribution assuming identical uncorrelated noise for both emitters such that the standard deviation of the distribution is given by $\sqrt{2}\sigma_\mathrm{noise}$. The result can be expressed using the scaled complementary error function $\mathrm{erfcx}(x)=e^{x^2}\mathrm{erfc}(x)$ as
\begin{equation}
    \langle{M_{12}\rangle}_\Delta = \frac{2\sqrt{\pi}}{\sigma_\mathrm{noise}\overline{\tau}}\mathrm{erfcx}\!\left(\frac{2}{\sigma_\mathrm{noise}\overline{\tau}M^\prime_{12}}\right),
\end{equation}
where $\overline{\tau}=(\tau_1+\tau_2)/2$ is the average lifetime of the emitters.
Note that $\lim_{x\rightarrow0}\mathrm{erfcx}(1/x)=x/\sqrt{\pi}$, and we recover $M_{12}^\prime$ in the limit $\sigma_\mathrm{noise}\rightarrow 0$.
The value of $\sigma_\mathrm{noise}$ is found to be \qty{4.8}{\percent} of the natural linewidth by fitting the RF intensity histogram (see Fig.\,\ref{fig1}e). Setting $\tau_1=\tau_2=\qty{200}{\ps}$ and $M_1=M_2=\qty{91}{\percent}$, $M_{12}$ decreases from \qty{91}{\percent} to \qty{90.7}{\percent} due to this level of spectral fluctuations, indicating a negligible impact of charge noise, within the experimental uncertainty ($\approx\SI{1}{\percent}$). For two ideal emitters, with $M_1=M_2=\qty{100}{\percent}$, this amount of charge noise provides an upper bound of $M_{12}=\qty{99.5}{\percent}$.\\ \\
\textbf{Photon indistinguishability measurement for a single-source.} For a single-source HOM interference measurement, the photons are sent into an unbalanced Mach-Zehnder interferometer (MZI). The first beam-splitter splits the photon stream into two arms, with one arm delayed by \qty{12.6}{\ns} to match the laser repetition rate (\qty{79.3}{\mega\hertz}). Two consecutively emitted photons interfere on the second beam-splitter and are detected with two avalanche photodiode single-photon detectors (APDs). From the detection events, two-photon coincidences are then analysed using a low-jitter (\qty{8}{\ps}) correlator (\textit{TimeTagger Ultra, Swabian Instruments}) with a chosen bin-size of \qty{50}{\ps}. To compute the single-source HOM visibility, we compare the area of the central peak ($A_\mathrm{central}$) with the mean area of the side-peaks ($\bar{A}_\mathrm{side}$), excluding the first peaks on either side of central peak. The HOM visibility is defined as
\begin{equation}
    V_\mathrm{HOM}=1-2\frac{A_\mathrm{central}}{\bar{A}_\mathrm{side}}\,.
\end{equation}
The correlation events within each peak are integrated over a \qty{4}{\ns} time window, which corresponds to approximately $\qty{20}{}$ times the emitter’s lifetime. For the computation of $g^{(2)}(0)$, we also use a time window of \qty{4}{\ns} for the central peak integration.
We extract the photon indistinguishability from the raw $V_\mathrm{HOM}$ value by correcting for setup imperfections and the non-zero $g^{(2)}(0)$ using the formula \cite{ollivier_hong-ou-mandel_2021}:
\begin{equation}
    M=\frac{V_{HOM}+4RT(1+g^{(2)}(0))-1}{4RT(1-g^{(2)}(0))}\,,
\end{equation}
which accounts for the imbalance of the second beam-splitter in the MZI, which we measure to be $R=0.47$ and $T=0.53$. \\ \\ 
\textbf{Mutual photon indistinguishability measurement for remote sources.} Regarding the two-source HOM interference, photons from both sources are sent to a fibre beam splitter whose outputs are connected to two APDs. We attenuate the count rate of S1 to match the count rate of S2 (\qty{1.5}{\mega\hertz}). Correlations are recorded using a bin size of \qty{50}{\ps}. To compute the two-source HOM visibility $V_\textrm{rem}$, we compare the area of the central peak in the parallel ($A_{\parallel}$) and orthogonal ($A_{\perp}$) configurations, both normalized by the mean area of their side peaks, and we define 
\begin{equation}
    V_\textrm{rem}=1-\frac{A_{\parallel}}{A{\perp}}\,.
    \label{eq3}
\end{equation}
The integration time window is set to \qty{4}{\ns}. We obtain the mutual indistinguishability $M_{12}$ using the following equation \cite{zhai_quantum_2022}:
\begin{equation}
    M_{12}=\left(\frac{R^2+T^2}{2RT}\right)\left[1+\frac{1}{2}(g^{(2)}_{\mathrm{S}1}(0)+g^{(2)}_{\mathrm{S}2}(0))\right]V_\mathrm{rem}\,,
\end{equation}
which accounts for the imbalance of the beam-splitter ($R=0.46$ and $T=0.54$) and the non-zero $g^{(2)}(0)$. The $g^{(2)}(0)$ of both sources is measured just before the two-source HOM experiment by blocking one beam-splitter input at a time. 
\\ \\
\section*{Data and materials availability}
Data and materials availability. All data acquired
and used in this work is property of the Centre for
Nanoscience and Nanotechnology and is available upon
reasonable request to pascale.senellart-mardon@cnrs.fr
or thibaut.pollet@universite-paris-saclay.fr.\\
\section*{Acknowledgements}
The authors thank Rinaldo Trotta and Helio Huet for the fruitful  discussions.
This work was conducted within the research program of the QDLight joint laboratory (C2N/Quandela), partially funded by  the TUF-TOPIQC project part of the Trilateral Call for Quantum Innovation co-financed by Germany, the Netherlands and by the French National Quantum Strategy (France 2030) program; as well as by the PROQCIMA program within the French National Quantum Strategy (France 2030). This work was partially supported by the Paris Ile-de-France Région in the framework of DIM QUANTIP, the French Defense ministry - Agence de l'innovatio de défense, the European Union’s Horizon CL4 program under the grant agreement 101135288 for EPIQUE project, the Plan France 2030 through the projects ANR22-PETQ-0011, ANR-22-PETQ-0006, and ANR-22-PETQ-0013. This work was carried out within the C2N micro-nanotechnologies platforms and was partially supported by the RENATECH network and the General Council of Essonne.\\
\section*{Author contribution}
T.P. conducted the experimental investigation, data analysis, methodology, and visualization. V.G. and W.H. performed the in-situ lithography. Wafer growth was conducted by A.P. and M.M. and A.L. provided help and supervision. D.T.and N.M. performed the sample fabrication and pre-characterization. P.Step. and P.Stein supported the optical measurements. J.A.S. investigated the low electrical-noise environment. S.B. supervised the device design and fabrication. S.W. and S.M. developed the numerical simulations and the theoretical model. T.P., P.Stein, S.W., S.B., and P.Sen. wrote the paper with feedback from J.A.S. and N.M. All authors participated to scientific discussions. P.Sen. supervised the project.

\section*{Competing interest} 
P.Senellart is a scientific advisor and co-founder of the company Quandela. The other authors declare no competing interests.

\section*{}
\setcounter{figure}{0}
\renewcommand{\thefigure}{\arabic{figure}}
\renewcommand{\figurename}{Extended Data Fig.}

\begin{figure*}[ht]
    \includegraphics[width=17.2cm]{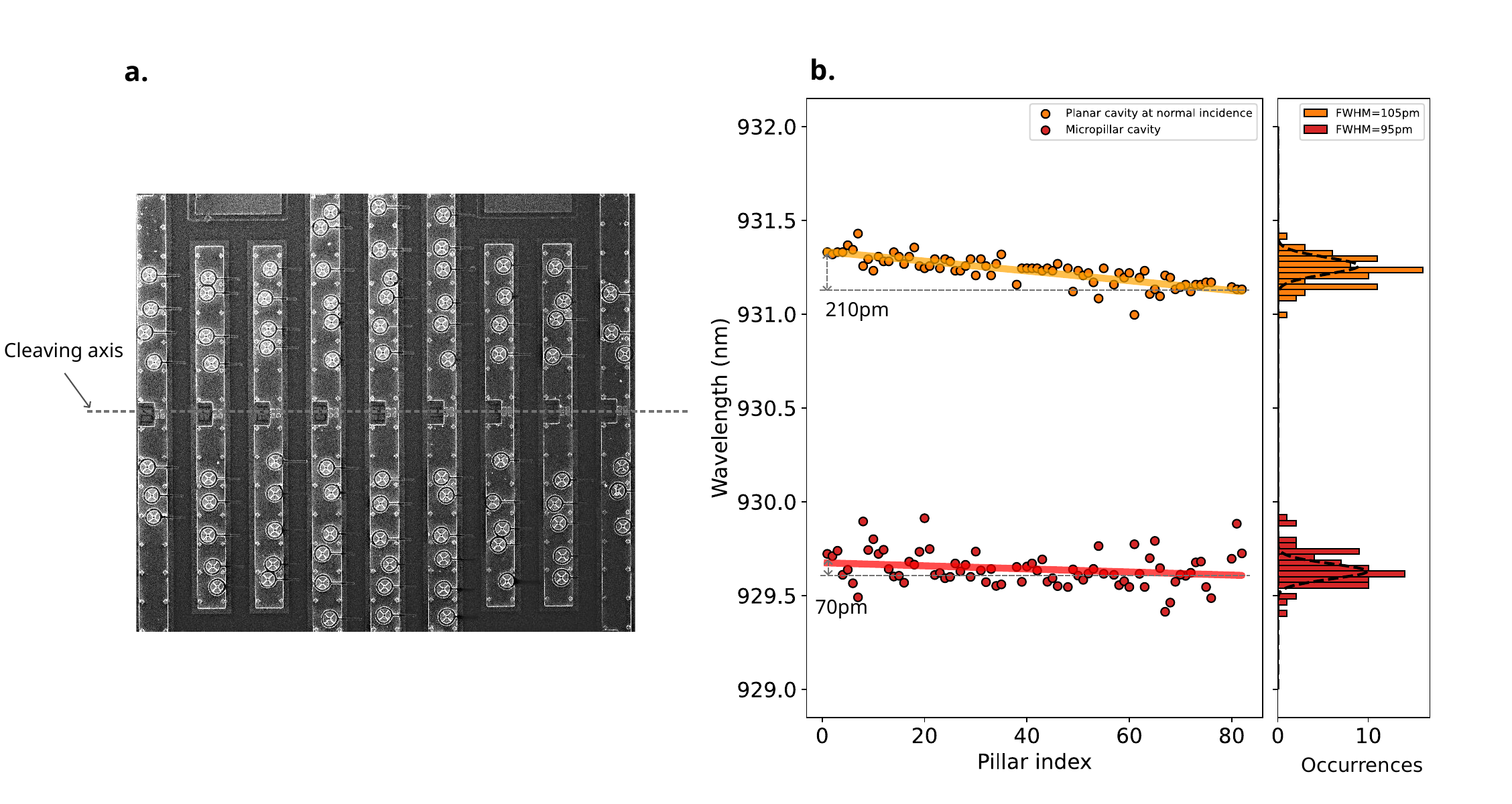}
    \caption{\label{extfig1}\textbf{Fabrication of identical micropillar cavities.} \textbf{a.} Scanning electron microscope image of the full sample. The sample is cleaved into two symmetrical pieces, each containing around \num{40} pillars. \textbf{b.} Planar and pillar cavity wavelengths as a function of the pillar indexes. Pillar indexes are ranked according to their distance from the center of the wafer. A linear fit reveals that the epitaxial thickness gradient of the wafer results in a shift of the planar cavity wavelength of \qty{210}{\pm}. This shift is reduced by a factor of three (\qty{70}{\pm}) by adjusting the diameter of each pillar during processing. The standard deviation of the micropillar wavelength distribution after etching is reduced with respect to the planar cavity distribution (from \qty{105}{\pm} to \qty{94}{\pm}), showing the high precision of our nanofabrication platform.}
\end{figure*}

\begin{figure*}[ht]
    \includegraphics[width=17cm]{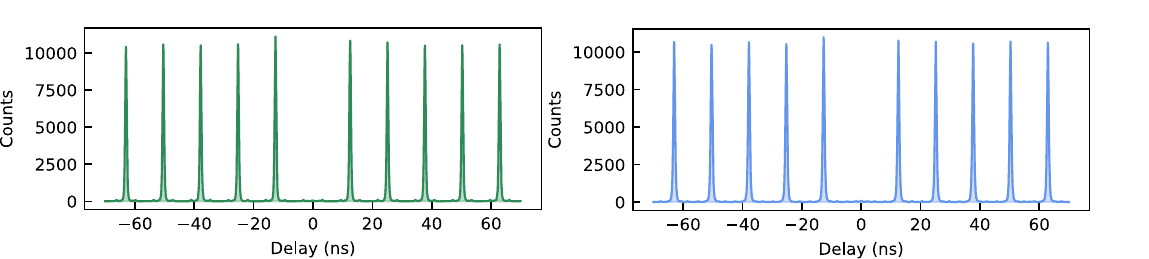}
    \caption{\label{extfig2}\textbf{Intensity autocorrelation measurements of S1 and S2.} Second-order autocorrelation of S1 and S2 obtained using LA excitation. S1 exhibits a $g^{(2)}(0)$ = \qty{1.4\pm.3}{\percent} and S2 a $g^{(2)}(0)$ = \qty{2\pm0.4}{\percent}.}
\end{figure*}

\begin{figure*}[ht]
    \includegraphics[width=15cm]{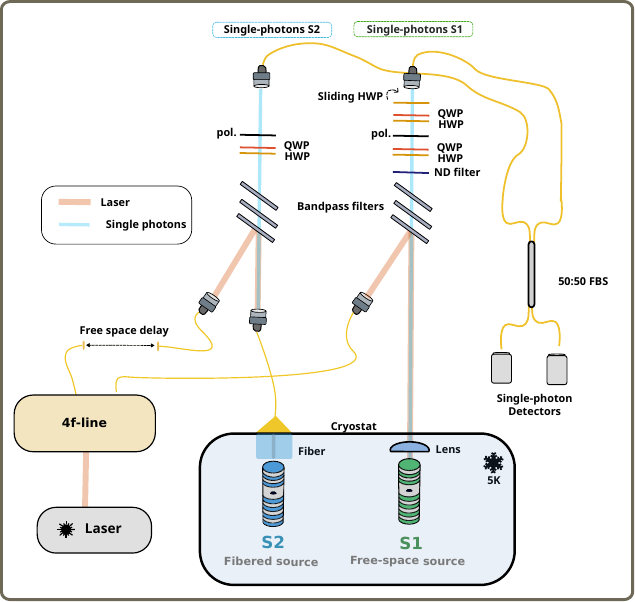}
    \caption{\label{extfig3} \textbf{Experimental setup.} The two sources are located in the same closed-cycle cryostat (\qty{5}{\kelvin}) on two different X,Y,Z nano-positioners stages.
    The source S1 is excited using a free-space optical setup, and a \num{0.7}\,NA lens is used to focus the laser on the micropillar. The source S2 is positioned under a high-NA fibre spliced to a single-mode fibre integrated inside the cryostat. A \qty{79.3}{\mega\hertz} repetition-rate laser is shaped with a 4-f shaping line to produce \qty{15}{\ps} long pulses. The laser light is then split into two arms and reflected on a bandpass filter (\qty{800}{pm} width) to S1 and S2, respectively.
    The reflected laser light is filtered using 3 bandpass filters and the polarized single photons from S1 and S2 are collected separately into two single-mode fibres.
    Both fibres are then connected to a \num{50}:\num{50} fibre beam-splitter, where the HOM interference occurs.
    Before collection, polarizers, half- and quarter-wave plates (HWP, QWP) are used to match the photon polarizations in both arms.
    An additional HWP mounted on a multi-position slider is added in one path to switch between the $\perp$ and $\parallel$ configuration. A free-space delay line with micrometer control is installed in one excitation path to match the arrival time of the photons on the FBS.
    Both outputs of the FBS are then connected to two fibrered single-photon avalanche photodiodes (APDs) with an efficiency of around \qty{31}{\percent} at \SI{929}{\nm}. 
    The two APDs are connected to a correlator with a low-time jitter, which is used to analyze two-photon coincidences.}
\end{figure*}

\begin{figure*}[ht]
    \includegraphics[width=15cm]{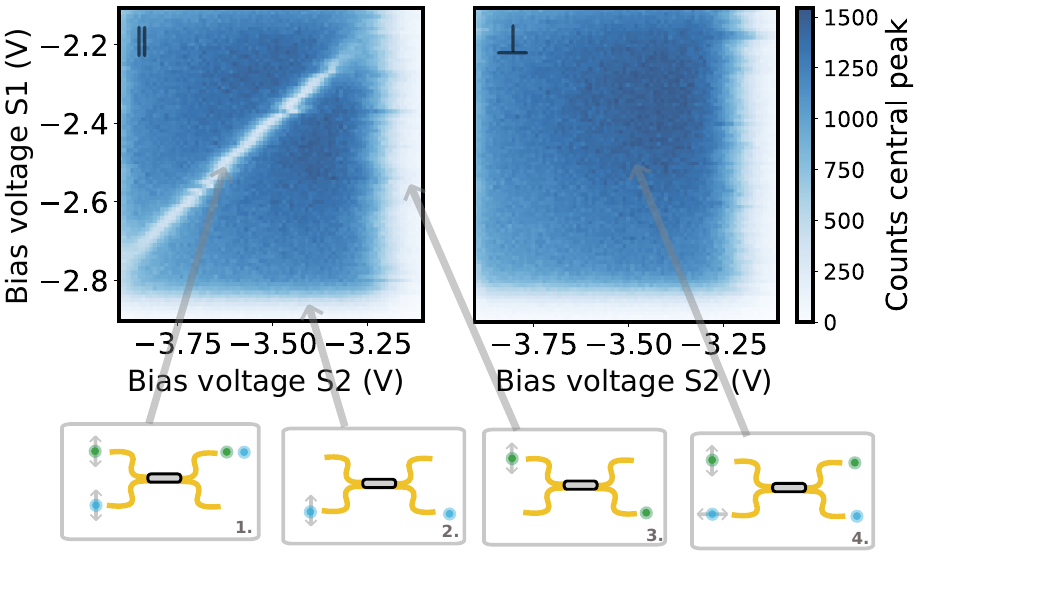}
    \caption{\label{extfig4} \textbf{Visualization of the HOM effect when the two sources are wavelength-matched.} Two-photon coincidences at zero time delay integrated in a \qty{4}{\ns} time window as a function of bias voltages applied to the sources in parallel ($A_{\parallel}$) and orthogonal ($A_{\perp}$) polarizations.
    This graph shows raw data ($A_{\parallel}$, $A_{\perp}$) used to calculate $V_\mathrm{rem}$. In contrast to the data shown in the main text (Fig.\,\ref{fig3}b), here we record the entire map in $\parallel$ before measuring in $\perp$, without the alternation for each data point. Additionally, we also scan across a wider voltage range, employ a lower integration time (\qty{0.1}{\s}) and use a different single-photon source compared to the one in the main text for S2. When the photons from S1 and S2 are co-polarized ($\parallel$), and at the same wavelengths, they are indistinguishable and interfere according to the HOM effect, leading to a reduction of $A_{\parallel}$.
    Consequently, a linear diagonal white region appears when there is no spectral detuning between S1 and S2 (1.). However, this region is missing when the photons are distinguishable (orthogonal in polarization, $\perp$), showing the absence of HOM interference (4.). The white regions on the side of the graph (2., 3.) demonstrate the single-photon purity from each source. When the applied bias voltage is too high, emission of the active trion state of one source is suppressed, and only photons from the other source reach the FBS, resulting in a simple intensity autocorrelation measurement ($g^2(\tau)$).}
\end{figure*}

\begin{figure*}[ht]
    \includegraphics[width=17.2cm]{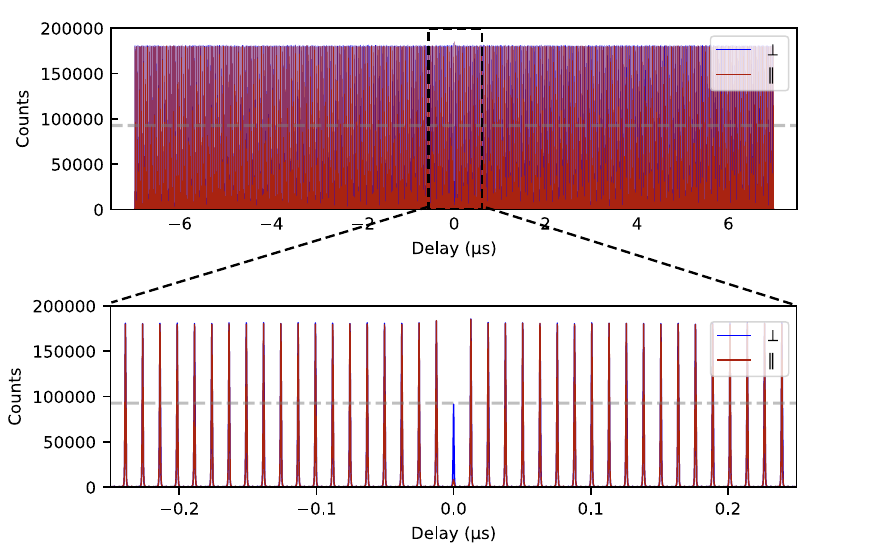}
    \caption{\label{extfig5} \textbf{HOM interference from remote QDs measured with longer delays.} HOM interference from photons emitted by two distant single-photon sources (S1 and S2), measured for delays up to \qty{7}{\micro\second} in parallel ($\parallel$) and orthogonal ($\perp$) polarizations. No blinking effect, which would result in an exponential decay of the peak envelope function, is observed on the \qty{}{\micro\second} timescale. The first two peaks (at \qty{-12.6}{\ns} and \qty{+12.6}{\ns}) are slightly higher than the others; an effect that we attribute to the spin dynamics of the individual negative trions, which is observed due to the random detection polarization basis with respect to the optical selection rules of the trion state. If the axis of the polarizer favors the transmission of one optical transition, more photons will be collected when the electron spin is coherent. This induces an increase of the two-photon coincidences when the spin is coherent, on a ns timescale, which impacts only the first two peaks of the correlation histogram. Note that this effect is also observed when measuring the autocorrelation function for individual sources and is not related to the HOM effect. In this paper, we did not use the area of these two peaks for any $g^{(2)}(0)$ or HOM measurement, as this would lead to a slight overestimation of the reported indistinguishability value ($ < \qty{0.03}{\percent}$).}
\end{figure*}

\end{document}